\documentclass[aip, apl, reprint, floatfix]{revtex4-1}
\usepackage{amsfonts, amssymb, amsbsy, amsmath}
\usepackage{graphicx}
\usepackage{hyperref}
\usepackage{bm}
\usepackage{dcolumn}

\begin{document}

\title{Laterally proximized aluminum tunnel junctions}

\newcommand{\mr}[1]{\mathrm{#1}}
\newcommand{\Vg}{$V_{\mr{g}}$}
\newcommand{\Vbias}{$V_{\textrm{bias}}$}
\newcommand{\Iset}{$I_{\textrm{SET}}$}
\newcommand{\Rt}{$R_{\textrm{T}}$}

\newcommand{\figta}{$\left(\mathrm{a}\right)\;$}
\newcommand{\figtb}{$\left(\mathrm{b}\right)\;$}
\newcommand{\figtc}{$\left(\mathrm{c}\right)\;$}
\newcommand{\figtd}{$\left(\mathrm{d}\right)\;$}
\newcommand{\figte}{$\left(\mathrm{e}\right)\;$}
\newcommand{\figtf}{$\left(\mathrm{f}\right)\;$}
\newcommand{\figa}{$\left(\mathrm{a}\right)$}
\newcommand{\figb}{$\left(\mathrm{b}\right)$}
\newcommand{\figc}{$\left(\mathrm{c}\right)$}
\newcommand{\figd}{$\left(\mathrm{d}\right)$}
\newcommand{\fige}{$\left(\mathrm{e}\right)$}
\newcommand{\figf}{$\left(\mathrm{f}\right)$}

\author{J. V. Koski}
\affiliation{Low Temperature Laboratory, Aalto University, POB 13500, FI-00076
AALTO, Finland} 

\author{J. T. Peltonen}
\affiliation{Low Temperature Laboratory, Aalto University, POB 13500, FI-00076
AALTO, Finland} 

\author{M. Meschke}
\affiliation{Low Temperature Laboratory, Aalto University, POB 13500, FI-00076
AALTO, Finland} 

\author{J. P. Pekola}
\affiliation{Low Temperature Laboratory, Aalto University, POB 13500, FI-00076
AALTO, Finland} 

\begin{abstract}
This letter presents experiments on junctions fabricated by a new technique that enables the use of high quality aluminum oxide tunnel barriers with normal metal electrodes at low temperatures. Inverse proximity effect is applied to diminish the superconductivity of an aluminum dot through a clean lateral connection to a normal metal electrode. To demonstrate the effectiveness of this method, fully normal-state single electron transistors (SET) and normal metal-insulator-superconductor (NIS) junctions applying proximized Al junctions were fabricated. The transport characteristics of the junctions were similar to those obtained from standard theoretical models of regular SETs and NIS junctions.
\end{abstract}

\date{\today}

\maketitle

An important element in nanoelectronic circuits is a tunnel barrier between two leads, with applications ranging from superconducting qubits~\cite{makhlin01} and quantum metrology devices~\cite{pekola08} to electronic refrigeration by hybrid normal metal-insulator-superconductor (NIS) junctions~\cite{giazotto06,nahum94,leivo96,clark04}. A standard material for yielding high quality junctions is aluminum due to rapid oxidation of its surface into a thin and stable layer of insulating aluminum oxide (AlOx) under oxygen exposure. Aluminum is superconducting below the critical temperature $T_{\rm{C}} =$ 1.2 K in bulk and up to 2.7 K in thin films~\cite{adams04}. High-quality nano-scale junctions are typically fabricated in a single vacuum cycle by shadow evaporation of metals at multiple angles through a suspended mask~\cite{dolan77}, combined with {\it in situ} oxidation. Depositing a layer of normal or superconducting material on top of oxidized Al forms a NIS or a superconductor-insulator-superconductor (SIS) junction. 

Regrettably, fabrication of low temperature normal metal-insulator-normal metal (NIN) junctions or NIS junctions with non-Al superconductor is not straightforward. The latter would be beneficial in terms of controlling the $T_{\rm{C}}$ through the choice of material: a higher $T_{\rm{C}}$ would benefit for example a current standard based on the hybrid SINIS turnstile~\cite{pekola08}, while a lower $T_{\rm{C}}$ would be efficient for low temperature NIS coolers~\cite{giazotto06}, where the optimal cooling power is at $T \simeq 0.7~T_{\rm{C}}$. In order to use Al for either NIN  junctions or NIS junctions with non-Al superconductor, its superconductivity must be suppressed. Alternatively, aluminum can be isotropically deposited as a thin layer on top of the N electrode, after which it is thoroughly oxidized before depositing the counterelectrode~\cite{gueron96}. When applied with shadow evaporation the latter technique, however, requires caution to ensure uniform Al coverage to avoid short circuits.

\begin{figure}[htb] 
\includegraphics[width = 1\columnwidth]{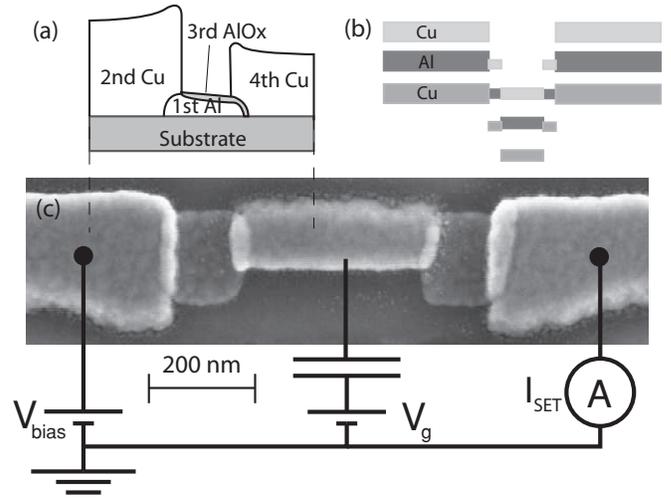}
\caption{\figta Sketch of a cross section view and the fabrication order of the NIN junction with copper leads. The aluminum is proximized via a clean contact to the left copper electrode. \figtb Sketch of the NININ SET shadow mask layout, including the positions of evaporated metals at different angles. \figtc Scanning electron micrograph (SEM) of a normal-state SET sample, integrated to the measurement circuit diagram.}
\label{fig:sample}
\end{figure}

The aluminum superconductivity may be suppressed by various means, such as by applying a magnetic field. However, magnetic fields are not always desirable when combining NIN and NIS junctions on a single circuit, for example when combining normal-state SETs with NIS thermometers and coolers~\cite{kubala08}. Although it is possible to have both NIN and NIS Al-based junctions in a single circuit by controlling the thicknesses and thus the critical fields of Al layers~\cite{greibe11}, in some applications external fields cannot be tolerated. Another approach is to dope Al with e.g. manganese impurities to control or diminish the $T_{\rm{C}}$ while retaining the BCS density of states~\cite{clark04, oneil08a}, but the junction quality could be compromised~\cite{oneil10}. Also the use of inverse proximity effect arising in a normal metal-superconductor (NS) bilayer is a widely used technique for modifying the superconductor $T_{\rm{C}}$~\cite{buzdin05,gupta04}: the normal metal gains a superconducting character, while the superconductor gains normal metal features at a short distance from the interface~\cite{buzdin05,pannetier00}.

This letter introduces a new technique for suppressing the Al superconductivity in small tunnel junctions, by utilizing the inverse proximity effect in lateral direction. According to the theory of diffusive, inhomogeneous proximity superconductivity~\cite{belzig99}, the superconductivity is fully suppressed in a NSN structure with infinite-size N reservoirs, when the length of the quasi-1D S-material wire $L < \pi \xi_0$ ~\cite{radovic91, liniger93}. By symmetry, for a NS structure the suppression is achieved for $L < \pi \xi_0 / 2$.  Utilizing our fabrication process the Al film superconducting coherence length is typically $\xi_0 \simeq 100$ nm~\cite{courtois08, peltonen10}, yielding $L \simeq 150$~nm. When accounting the non-ideal interface, a short Al wire with $L \sim 100$~nm, proximized by a long normal metal lead, forms a high-quality NIN or NIS tunnel junction.

The proximized Al junctions were utilized in the fabrication of several normal-state SET samples with multi-angle shadow evaporation, for which the mask layout is shown in Fig.~\ref{fig:sample}~\figb. Figure~\ref{fig:sample}~\figta sketches a cross-section of a proximized Al junction that is used in the sample. The fabrication process had the following order, illustrated in Fig.~\ref{fig:sample}~\figa: first a 15 nm layer of aluminum is evaporated to form two small Al dots 400 nm apart, directly followed by a thick (45 nm) copper in a different angle to partially cover the dot with a proximizing normal metal lead. Third, the remaining uncovered aluminum is oxidized in the evaporation chamber, after which the SET is formed by depositing a 30 nm copper layer at a third angle to contact the two Al dots. 

\begin{figure}[ht]
\includegraphics[width =  1\columnwidth]{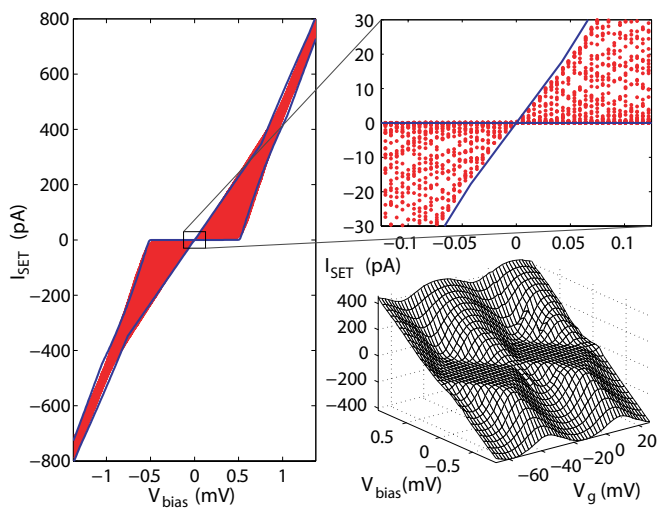}
\caption{ (color online) \figta Measured current \Iset~through the NININ SET with normal state resistance $R = R_1 + R_2 \simeq 1.2$~M$\Omega$ and charging energy at $T = 63$~mK. The gate voltage $V_g$ was repeatedly swept over several gate periods while the bias voltage \Vbias~was increased slowly. The blue lines are theoretical values for normal SET minimum and maximum currents. \figtb An enlarged view of \figa. The maximum of \Iset~has a linear dependence on \Vbias~around \Vbias= 0. \figtc \Iset~as a function of \Vbias~and $V_g$. The coulomb diamonds of a fully normal SET are clearly visible. }
\label{fig:currentbytime}
\end{figure}

The suppression of the Al dot superconductivity was examined by measuring the current \Iset~as a function of applied bias and gate voltages, \Vbias~and $V_g$, respectively, according to the circuit diagram in Fig.~\ref{fig:sample}~\figc. The samples were measured in a $^3$He -$^4$He dilution refrigerator~\cite{pekola94} with a base temperature close to 40 mK. The results for the SET of Fig.~\ref{fig:sample}~\figtc with highest charging energy $E_C \simeq 525$~$\mu$eV are shown in Fig.~\ref{fig:currentbytime}.  All of the measured structures showed similar behavior. Figures \figta and \figtb show the range over which \Iset~varies as $V_g$ is swept, fit by a standard SET model curve based on the orthodox theory\cite{averin86}, with the inclusion of asymmetric junction resistances, $R_1 \simeq 0.3~R_2$. The linearity at low bias voltages for maximum current indicates that the leads in the SET are normal, for otherwise there would be a gate-independent zero-current region for $|e$\Vbias$| < \Delta$.\cite{pekola08} This is further supported by Fig.~\ref{fig:currentbytime}~\figc, which shows \Iset~as a function of both \Vbias~and $V_g$, demonstrating the Coulomb diamonds characteristic to a normal SET.

\begin{figure}[ht]
\includegraphics[width =  1\columnwidth]{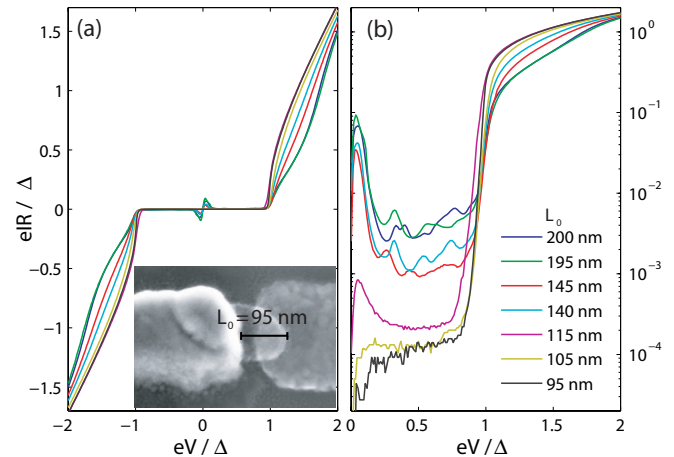}
\caption{(color online) IV characteristics of several of the measured samples with proximized Al - AlOx - Al NIS junctions on \figta linear and \figtb semilogarithmic scale. A clear sign of supercurrent is visible in the samples with longest Al dots. The lengths $L_0$ are indicated in the figure. As $L_0$ decreases, the supercurrent peak gradually diminishes with only a minor zero-voltage step remaining for the junctions with the smallest $L_0$. Inset: SEM micrograph of the sample with the shortest Al dot, indicating the length $L_0 \simeq 95$~nm. The copper lead to the left proximizes the Al dot in the center, which is connected to the Al lead to the right via a tunnel barrier. Due to the limitations in the mask design, the 20~nm thick Al lead was fabricated first and oxidized, before depositing the Al dot on top. However, the Al is proximized in the same way as previously. 
}
\label{fig:NISCurrent}
\end{figure}

\begin{figure}[tb]
\includegraphics[width =  1\columnwidth]{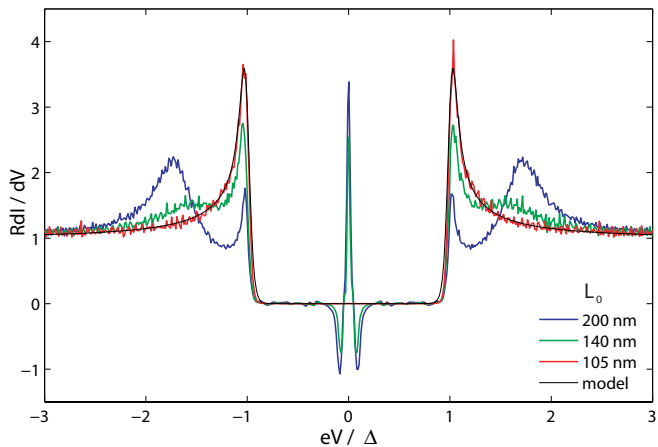}
\caption{(color online) Differential conductance of some proximized Al NIS junctions as a function of voltage, obtained by numerical differentiation of the IV characteristics. A fit based on the standard theoretical model for a NIS junction is drawn in black. A transition from SIS-type to a NIS-type $dI/dV$ dependence occurs with decreasing $L_0$. The samples with the shortest proximized Al dots behave as standard NIS junctions. }
\label{fig:NISDiffCurrent}
\end{figure}

Another more demanding test of the quality of a proximized junction is to utilize them in a NIS junction. The magnitude of subgap features, such as supercurrent through the junction, is an excellent indicator for determining the superconductivity of the Al dot. Also the differential conductance of the junction at low $T$ is directly proportional to the superconductor density of states if the counterelectrode is at normal state. The NIS junction samples are fabricated like the NIN junctions of the SET described earlier, with Al as the S lead and copper as the N lead, yet these could be replaced by other normal and superconducting metals. 

\begin{figure}[htb]
\includegraphics[width =  1\columnwidth]{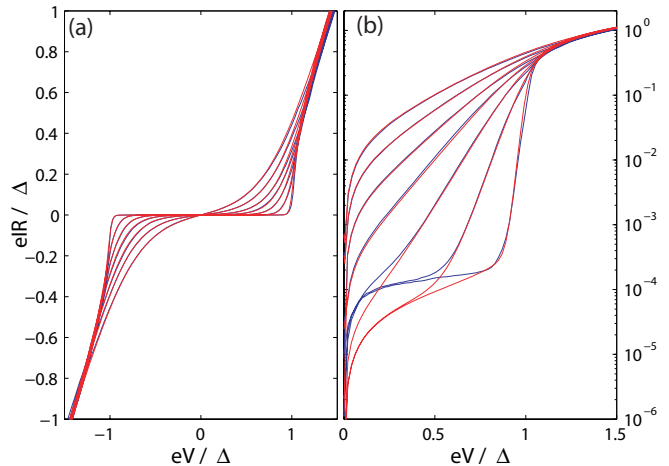}
\caption{(color online) IV characteristics of the NIS junction with $L_0 = 95$~nm at various temperatures (solid blue lines), together with a theoretical fit of the current (solid red lines) on linear \figta and semilogarithmic \figtb scale. There is only a minor deviation at low $V$ and $T$ where the Dynes parameter does not explain the current step-formation at low $V$ and $T$, however otherwise the IV characteristics are very similar to standard NIS junctions.}
\label{fig:NISCurrentHeating}
\end{figure}

Several NIS junctions were fabricated simultaneously with various $L_0$, the length of Al dot uncovered by the normal electrode. The NS contact and tunnel junction overlap areas were kept constant. The inset of Fig.~\ref{fig:NISCurrent}~\figta shows a SEM image of the structure with the shortest $L_0 \simeq 95$ nm. The measured normal state resistances $R$ of the junctions were in the range of $35 - 75$~k$\Omega$ and the zero temperature superconductor energy gap of the S - Al was $\Delta \simeq 220~\mu$eV. Figure ~\ref{fig:NISCurrent} shows the measured current-voltage (IV) characteristics at $T = 65$~mK. Additionally, Fig.~\ref{fig:NISDiffCurrent} shows the differential conductance of several samples obtained by numerical differentiation of the IV data together with a theoretical fit for a standard NIS junction. Finally, the IV characteristics of the $L_0 \simeq 95$~nm sample in the temperature range of $T = 43 - 675$~mK are plotted in Fig.~\ref{fig:NISCurrentHeating}, together with theoretical curves using a phenomenological Dynes parameter~\cite{dynes84} $\gamma \simeq 1.5 \times 10^{-4}$ in the Al quasiparticle density of states.

In these figures the dependence on the length of the Al dot is clearly visible: with $L_0 \simeq 200$ nm they appear to have mixed NIS and SIS junction properties with notable supercurrent peaks, but as the dot length approaches 100 nm, both the IV curve and the differential conductance resemble increasingly those of a regular NIS junction. The results suggest that proximized Al tunnel junctions provide a valid tool for fabricating NIS junctions, where the superconductor is not necessarily aluminum. Such structures have been a challenge to fabricate by shadow evaporation till now.

Concluding, the method of forming tunnel junctions with the help of  laterally proximized Al was introduced. These junctions were implemented in a fully normal SET and individual NIS junction configurations. The measured IV characteristics would indicate that the presented method is likely to allow shadow fabrication of high quality NIN junctions, as well as NIS junctions with non-Al superconductor. 

We acknowledge financial support from the European Community's FP7 Programme under Grant Agreement No. 228464 (MICROKELVIN, Capacities Specific Programme) and No. 218783 (SCOPE). We thank T. Heikkil\"a and P. Virtanen for discussions.

\end{document}